\documentclass[11pt,a4paper]{article}
\pdfoutput=1 
\usepackage{jheppub} 
\usepackage[utf8]{inputenc}
\usepackage{amsfonts}
\usepackage{amsbsy}
\usepackage{url}
\usepackage{enumerate}
\usepackage[justification=raggedright]{caption}
\usepackage{hhline}
\usepackage{graphicx,color}
\usepackage{amssymb,amsmath}
\usepackage{slashed}
\usepackage{hyperref}
\usepackage{braket}
\usepackage{simplewick}
\usepackage{ascmac}
\usepackage{latexsym}
\usepackage{pifont}          
\usepackage{bm}
\usepackage{comment}
\usepackage[normalem]{ulem}
\usepackage{CJKutf8}
\usepackage{cancel}
\graphicspath{{./fig/}{./}{./figs/}}
\usepackage[dvipsnames]{xcolor}

\begin{document}
\title{Baryonic vortices in rotating nuclear matter}

\author[a,b]{Kazuya~Mameda,}
\emailAdd{k.mameda@rs.tus.ac.jp}
\affiliation[a]{Department of Physics, Tokyo University of Science, Tokyo 162-8601, Japan}
\affiliation[b]{RIKEN iTHEMS, RIKEN, Wako, Saitama 351-0198, Japan}
\author[c,d]{Muneto~Nitta}
\emailAdd{mune.nitta@gmail.com}
\affiliation[c]{Department of Physics, and 
Research and Education Center for Natural Sciences, Keio University, 4-1-1 Hiyoshi, Yokohama, Kanagawa 223-8521, Japan}
\affiliation[d]{International Institute for Sustainability with Knotted Chiral Meta Matter(WPI-SKCM$^2$), Hiroshima
University, 1-3-2 Kagamiyama, Higashi-Hiroshima, Hiroshima 739-8511, Japan}

\author[c]{and Zebin~Qiu}
\emailAdd{qiuzebin@keio.jp}

\preprint{RIKEN-iTHEMS-Report-26}

\abstract{
We investigate baryonic vortices as topological excitations in rotating nuclear matter within the framework of chiral perturbation theory. 
We identify two distinct configurations: local and global vortices, both carrying the baryon number as the topological charge associated with the third homotopy group $\pi_3(S^3)$.
For the local vortex, similar to the vortex Skyrmion in a finite isospin chemical potential, charged pions form the condensate on the boundary and have a phase winding, while the neutral pion varies along the rotation axis inside the vortex core.
On the other hand, a global vortex is formed by the condensate and phase winding of the neutral pion, while the charged pions vary on the inside along the rotation axis.
Crucially, although global vortices are usually discarded in infinite systems due to logarithmic divergence in energy, we demonstrate that the finite-size constraint dictated by causality in a rotating frame regularizes the divergence physically, rendering the global vortex a viable excitation.
We reveal an energetic competition between global and local vortex states, under the tunable parameters of rotation, system size, and baryon chemical potential. 
Our results suggest that the previously overlooked global vortex can play a significant role in the topological structure of rotating dense QCD matter.
}

\maketitle

\section{Introduction}
\label{sec:intro}

The phase structure of Quantum Chromodynamics (QCD) under extreme environments has been a subject of intense study, revealing a rich landscape of physical phenomena beyond the perturbative regime. 
In particular, triggered by the observation of the ``most vortical fluid'' of quark-gluon-plasma (QGP) in heavy-ion collision experiments~\cite{STAR:2017ckg}, the response of QCD matter to rapid rotation, such as chiral symmetry breaking and confinement has been extensively investigated through various approaches~\cite{Chen:2015hfc,Jiang:2016wvv,Chernodub:2016kxh,Chen:2022mhf,Chen:2023cjt,Fujimoto:2021xix,Chernodub:2020qah,Braga:2022yfe,Chen:2020ath,Mameda:2023sst,Kawaguchi:2025mkh}.
On top of these works, which are primarily focused on quarks and gluons, the impact of external rotation is also found in the hadronic phase.
Significant progress has been made in understanding how rotation couples to hadronic matter via the chiral anomaly.
For instance, the role of the hadronic vertex required by chiral transport under rotation has been analyzed~\cite{Huang:2017pqe,Nishimura:2020odq}, 
the ground state with $\eta(')$ chiral soliton lattice 
\cite{Nishimura:2020odq,Eto:2021gyy,Eto:2023tuu,Eto:2023rzd} stabilized by 
the Wess-Zumino-Witten (WZW) term \cite{Witten:1983tw}
has been proposed,
and hadronic transport phenomena induced by the WZW term coupled with a vortical field~\cite{Evans:2025jsa}
have been explored.

A central question that naturally arises in this context is the effect of rotation on the topological excitations of baryons, specifically the 
baryonic vortex, which is not fully addressed by previous analyses of rotating baryonic matter without dynamical electromagnetic fields.
In the presence of finite density and magnetic fields, baryons can emerge as topological solitons trapped within vortices. Such a baryonic vortex lattice can be the ground state of dense QCD matter in strong magnetic fields~\cite{Qiu:2024zpg,Hamada:2025inf,Hamada:2026uec}.
Although the well-known classical correspondence between magnetism and rotation invokes analogous structures stabilized or modified by a mechanical rotation, the phase diagram involves involving baryonic vortices is still under examination. 
Solving this problem requires a careful treatment of the dynamics of electromagnetic fields coupled with rotation.
If a baryonic vortex formation dynamically induces a magnetic field, the system becomes an environment of coexisting rotation and magnetism, where mechanical stabilization becomes more nontrivial~\cite{Fukushima:2024tkz}.

A typical feature of relativistic rotating systems is the finite-size effect required by causality, which is not merely a technical artifact but has profound implications for infrared dynamics~\cite{Ebihara:2016fwa,Kuboniwa:2025vpg} and the stability of topological defects.
While in infinite systems, global vortices associated with the winding of the neutral pion field are typically discarded due to their logarithmically divergent energy, in a finite-size rotating system, this divergence is regularized, potentially rendering the global vortex a physical solution that competes with a local (gauged) vortex.

In this paper, we investigate the formation and stability of baryonic vortices in rotating nuclear matter. 
We employ chiral perturbation theory (ChPT) up to $\mathcal{O}(p^2/\Lambda_\chi^2)$ coupled with dynamical electromagnetic fields, explicitly incorporating the WZW term, which is essential for the baryonic nature of the vortex. 
We focus on the massless limit to elucidate the topological and finite-size effects driven by rotation, and examine whether local and global vortices manifest as stable excitations or ground states within the causal boundary of the rotating system.
Whereas realistic heavy-ion collisions involve high temperatures, investigating the zero-temperature limit is a crucial first step. It allows us to purely extract the topological consequences of the WZW term and the finite-size effects driven by causality, providing a rigorous theoretical baseline for topological excitations in rotating baryonic matter before thermal fluctuations are introduced.

Let us mention some related works. 
Non-topological pion strings have been studied in Refs.~\cite{Zhang:1997is,Berera:2016vhw} in which the stabilization by thermal effects is discussed.
However, the chiral anomaly was not taken into account.
In our study, the chiral-anomaly induced WZW term lowers the energy of pion vortex strings, 
enhancing their stability. 
In the context of ChPT for low-energy QCD,
local vortices, {\it i.e.},
Abrikosov–Nielsen–Olesen (ANO) vortices~\cite{Abrikosov:1956sx,Nielsen:1973cs} have been explored 
at finite baryon or isospin chemical potential~\cite{Adhikari:2015wva,Adhikari:2018fwm,Adhikari:2022cks,Gronli:2022cri,
Canfora:2020uwf,Canfora:2024mkp,Evans:2022hwr,Evans:2023hms}.

This paper is organized as follows.
In sec.~\ref{sec:CPT}, we explain our scenario and prescribe the chiral Lagrangian with WZW term under rotation.
In sec.~\ref{sec:local vortex}/sec.~\ref{sec:global vortex}, we discuss local/global vortices as solutions to the theory.
In sec.~\ref{sec:phase}, we investigate a transition between two types of vortex states and discuss its relevance to QGP and neutron stars. 
Section \ref{sec:summary} is for summary and discussion.


\section{Chiral Lagrangian under Rotation}
\label{sec:CPT}

Our intuition is that a rotating system may accommodate vortices, e.g., superfluid vortices. For a dense baryonic system, a vortex could become the ground state if it carries baryon number, because then it couples with the baryon chemical potential to reduce the free energy via the WZW term. 
A typical way to realize a baryon number in a vortex state is by having a neutral (charged) pion field that varies along the vortex worldline, in the vortex core of charged (neutral) pions. Such a configuration carries a nontrivial topological charge of $\pi_3(S^3)\simeq {\mathbb Z}$, identified as a baryon number.\footnote{Such configurations can be stable even without a Skyrme term, unlike the conventional Skyrmions \cite{Skyrme:1961vq,Skyrme:1962vh}.
This is because the presence of a host soliton allows a Skyrmion in its worldvolume.
Another example is a domain-wall Skyrmion~\cite{Nitta:2012wi,Nitta:2012rq,Gudnason:2014hsa,Gudnason:2014nba,Eto:2015uqa}, namely a Skyrmion living inside a pion domain wall. 
Under a strong magnetic field at high density, there exists a domain-wall Skyrmion phase~\cite{Eto:2023lyo,Eto:2023wul,Amari:2024fbo,Eto:2023tuu,Amari:2024mip,Amari:2025twm,Copinger:2025rpo}.
}
We refer to such a state
as the vortex-Skyrmion~\cite{Gudnason:2014hsa,Gudnason:2014jga,Gudnason:2016yix,Nitta:2015tua,Nitta:2022ahj} (regarding the form of the ansatz), or the baryonic vortex~\cite{Qiu:2024zpg,Hamada:2025inf,Hamada:2026uec} 
 (regarding its physical nature as an effective description
of baryons). The baryonic vortex sustains a magnetic field via quantized
magnetic flux as a type-II superconductor. At high enough baryon density,
the magnetic field can be induced spontaneously, which may inspire scenarios for neutron star cores.

As a first step, we consider a steady mechanical rotation along a
certain axis $\boldsymbol{\Omega}=\Omega\hat{z}$.
Then a natural conjecture in view of symmetry is that if a generated magnetic field emerges, $\boldsymbol{B}=B\hat{z}$ along $\hat{z}$ should be at the lowest cost of energy. 
For simplicity, a $U\left(1\right)$ gauge field $A=A_\phi\left(\rho\right)d\phi$, 
spanned on cylindrical coordinates $(t,\rho,\phi,z)$, suffices to produce the $\boldsymbol{B}$ that we consider. 
Later on, we will specify how it is compatible with our soliton. 
We study the phenomenon in a co-moving frame, where the rotational effects are encoded in the metric
\begin{equation}
g_{\mu\nu}=
\begin{bmatrix}
1-\rho^{2}\Omega^{2} & 0 & -\rho^{2}\Omega & 0\\
0 & -1 & 0 & 0\\
-\rho^{2}\Omega & 0 & -\rho^{2} & 0\\
0 & 0 & 0 & -1
\end{bmatrix}.
\label{eq:gmunu}
\end{equation}
The soliton is static in such a frame. An important feature here is that we need to take into account the causality constraint and thus a finite system size:
\begin{equation}
\label{eq:causality}
 0\leq \rho \leq R \leq \Omega^{-1},
\end{equation}
where $R$ is the transverse ($xy$-plane) radius of the system.

Our targeted regime is the low-energy QCD described by ChPT. It is an effective theory based on the momentum expansion of pion degrees of freedom, specified by a power counting 
$\mathcal{O}\left(p^n/\Lambda_\chi^n\right)$ with the UV cutoff $\Lambda_\chi = 4\pi f_\pi/\sqrt{N_\mathrm{f}}$, where $f_\pi$ is the pion decay constant~\cite{Manohar:1983md}.
For simplicity, we consider the chiral limit. The leading order Chiral Lagrangian for $N_\mathrm{f}=2$ yields the kinetic term:
\begin{equation}
S_{\text{kin}}= \frac{f_{\pi}^{2}}{2} \int d^4x \sqrt{-\det g_{\mu\nu}}\, g^{\mu\nu}\mathrm{Tr}\left(D_{\mu}U^{\dagger}D_{\nu}U\right),
\end{equation}
gauged by the electromagnetic $U\left(1\right)$
with covariant derivative
\begin{equation}
D_{\mu}U=\partial_{\mu}U-iA_{\mu}\left[Q,U\right].\label{eq:D}
\end{equation}
The electric charge $e$ is absorbed into the definition of $A_{\mu}$ and $Q$ is the electric charge matrix amounts to $\mathbb{I}/6+\tau^{3}/2$ with $\tau^{i}$ the Pauli matrices.  
The meson field $U\in SU(2)$ can be written as 
\begin{equation}
U=\sigma+i\boldsymbol{\tau}\cdot\boldsymbol{\pi}.
\end{equation}
The stress tensor is defined by the variation with respect to the metric. For instance, the kinetic term contributes
\begin{equation}
T^{\mu\nu}_\text{kin}=-\frac{2}{\sqrt{-\det g_{\mu\nu}}}\frac{\delta S_\text{kin}}{\delta g_{\mu\nu}}.
\label{eq:Tdef}
\end{equation} 
For the $g_{\mu\nu}$ in eq.~\eqref{eq:gmunu} one can prove $\sqrt{-\det g_{\mu\nu}}=\rho$ and hence  
\begin{equation}
T_{\text{kin}}^{\mu\nu}=\frac{f_{\pi}^{2}}{4}\left(2g^{\mu\lambda}g^{\nu\kappa}-g^{\mu\nu}g^{\lambda\kappa}\right)\mathrm{Tr}\left(D_{\lambda}U^{\dagger}D_{\kappa}U\right).
\end{equation}
This leading order $\mathcal{O}\left(p^{2}/\Lambda_\chi^2\right)$ stress tensor is enough to explain our vortex ansatz. The
unitarity $U^{\dagger}U=1$ yields the parametrization 
\begin{equation}
\begin{split}
\sigma=\cos\alpha\cos\beta, & \quad\pi_{3}=\cos\alpha\sin\beta, \\
\pi_{1}=\sin\alpha\sin\gamma, & \quad\pi_{2}=\sin\alpha\cos\gamma,  
\end{split}
\end{equation}
among which $\alpha$, $\beta$ and $\gamma$ are functions of spatial
coordinates $\left(\rho,\phi,z\right)$. 
According to the rotational symmetry brought by $\boldsymbol{\Omega}$, either $\beta$ or $\gamma$
can be identified with the azimuthal angle $\phi$ in view of minimizing the energy. 
The kinetic contribution
to energy density $\mathcal{E}$ reads
\begin{equation}
\mathcal{E}_{\text{kin}}=T^{0\lambda}_\text{kin}g_{\lambda0}=\frac{f_{\pi}^{2}}{2}\left[
\left(\frac{\partial\sigma}{\partial\rho}\right)^2+\left(\frac{\partial\sigma}{\partial z}\right)^2+\left(\frac{\partial\boldsymbol{\pi}}{\partial\rho}\right)^2+\left(\frac{\partial\boldsymbol{\pi}}{\partial z}\right)^2
+\left(\frac{1}{\rho^{2}}-\Omega^{2}\right)\Xi\right],
\label{eq:Ekin}
\end{equation}
where $\Xi$ is the part involving $\phi$-derivatives and the gauge field $A_{\phi}$:
\begin{equation}
\Xi=
\begin{cases}
\left(\sigma^{2}+\pi_{3}^{2}\right)+\left(\pi_{1}^{2}+\pi_{2}^{2}\right)
A_{\phi}^{2};\quad\beta=\phi,\\[0.5em]
\left(\pi_{1}^{2}+\pi_{2}^{2}\right)\left(1-A_\phi\right)^{2};\quad\gamma=\phi.
\end{cases}
\label{eq:2kin}
\end{equation}
Note that if we had no dynamical gauge field, the two types of ansatz with $\beta=\phi$ and $\gamma=\phi$ would be degenerate as $\rho^{-2}-\Omega^{2}$ can couple with either $\sigma^{2}+\pi_{3}^{2}$ or $\pi_{1}^{2}+\pi_{2}^{2}$.
This is natural because rotation does not distinguish between charged and uncharged particles. Either ansatz specified in eq.~\eqref{eq:2kin} can be categorized as a vortex given the phase $\phi$.

The dynamics in our system are determined by not only the kinetic action but also the Maxwell action, which takes the following form in the general coordinates
\begin{equation}
S_{\text{EM}}=-
\frac{1}{4e^{2}} \int d^4x \sqrt{-\det g_{\mu\nu}} \,g^{\mu\rho}g^{\nu\sigma} F_{\mu\nu}F_{\rho\sigma},
\end{equation}
In the same way as eq.~\eqref{eq:Tdef}, we derive the electromagnetic sector of the stress tensor 
\begin{equation}
T_{\text{EM}}^{\mu\nu}=\frac{1}{e^{2}}\left(-g^{\mu\lambda}F_{\lambda\kappa}g^{\kappa\nu}+\frac{1}{4}g^{\mu\nu}F_{\lambda\kappa}F^{\kappa\lambda}\right),
\end{equation}
which leads to an energy density in our setup as
\begin{equation}
\mathcal{E}_{\text{EM}}=\frac{1}{2e^{2}}\left(\frac{1}{\rho^{2}}-\Omega^{2}\right)\left(\frac{\partial A_\phi}{\partial\rho}\right)^{2}.
\end{equation}

Apart from the Maxwell action, $A_\mu$ involves an anomaly effect, encoded in the WZW action.
Our highlight is a vortex-Skyrmion state of baryons, which can feature lower energy than pure pionic vortices in that it carries a baryon number coupling with
the baryon chemical potential $\mu$ to contribute a negative portion of the energy via the gauged WZW term:
\begin{equation}
S_{\text{WZW}}=\int d^4x \sqrt{-\det g_{\nu\lambda}} \, j_B^{\mu}\left(qA_{\mu}+A_{\mu}^B\right).
\end{equation}
Among it, $A_\mu^B=\left(\mu,0,0,0\right)$ is an effective baryon gauge field capturing the finite $\mu$ ~\cite{Son:2007ny}.
$j_B^\mu$ is the Goldstone-Wilczek current that can be written with differential forms as \cite{Goldstone:1981kk}
\begin{equation}
j_B=\star\frac{1}{24\pi^{2}}\mathrm{Tr}\left\{ l\wedge l\wedge l+3iQd\left[A\wedge\left(l-r\right)\right]\right\}.
\label{eq:jB}
\end{equation}
We use $l=U^{\dagger}dU$ and $r=UdU^{\dagger}$ but can equivalently rephrase $j_B^\mu$ using $D_\mu$ defined in eq.~\eqref{eq:D}, c.f. Refs.~\cite{Son:2007ny,Goldstone:1981kk}.
$j_B^{\mu}$ is physically interpreted as the baryon current. 
We further scrutinize the expression of the baryon density $j_B^{0}$ by applying both types of the vortex ansatz
bifurcated from eq.~\eqref{eq:2kin}.
\begin{equation}
j_B=\star\frac{1}{4\pi^{2}}\left\{ \begin{array}{c}
d\left(\cos^{2}\alpha\right)\wedge d\phi\wedge d\gamma;\quad\beta=\phi\\
-d\left[\cos^{2}\alpha\left(1-A_\phi\right)\right]\wedge d\phi\wedge d\beta;\quad\gamma=\phi
\end{array}\right..\label{eq:2jB}
\end{equation}
The expression shows that a nonvanishing baryon number $\int d^{3}xj_B^{0}=N$ cannot be realized if there are only charged pions. 
In other words, the neutral pion proves indispensable in endowing the baryonic nature to a vortex.
The energy density resulting from $S_{\text{WZW}}$ reads
\begin{equation}
\mathcal{E}_{\text{WZW}}=-\mu j^0_B.
\end{equation}
Its contribution to the total energy reads $-\mu N$, which is topological, {\it i.e.}, independent of $g_{\mu\nu}$.
The integer $N$ is the topological charge.

In all, what is explored in the current work is the baryonic vortex governed by the action 
$S=S_\text{kin}+S_\text{EM}+S_\text{WZW}$. 
In such a framework,
a vortex solution is achieved by the field configurations that minimize the total free energy 
\begin{equation}
\label{eq:totalE}
 E = \int d^3 x\mathcal{E},
 \quad
 \mathcal{E} =\mathcal{E}_\mathrm{kin} + \mathcal{E}_\mathrm{EM} + \mathcal{E}_\mathrm{WZW}.
\end{equation}
We remark as we did in preceding works~\cite{Qiu:2024zpg,Hamada:2025inf,Hamada:2026uec}: at sufficiently large $\mu$, the longitudinal ($z$-axis) scale of a vortex collapses to infinitesimal because we only take into account $\mathcal{O}\left(p^{2}/\Lambda_\chi^2\right)$
terms in the Lagrangian. 
The consequence is an infinitely large longitudinal
momentum that violates the validity of ChPT. 
One way to avoid such an artifact is to add an $\mathcal{O}\left(p^{4}/\Lambda_\chi^4\right)$ Skyrme term, which is known for overcoming the Derrick's scaling law and describing a baryon as a Skyrmion with finite size. 
Nevertheless, the Skyrme term depends on an extra parameter that could be tuned in varied ways, e.g., by fitting the experimental data of either hadron masses, or axial coupling constant, among others~\cite{Adkins:1983ya}. 
In order to keep our results model independent, we do not include the Skyrme term, restricting the study within leading order ChPT. 

Before presenting the concrete solutions, let us first give an intuitive picture of both types of the vortices, corresponding to $\gamma=\phi$ and $\beta = \phi$ in eqs.~\eqref{eq:2kin}.
The former yields a gauged local vortex with $A_\phi = 1$ at $\rho \to R$ for $\Xi$ to vanish on the boundary, which is favored due to the positivity of $\left(R^{-2}-\Omega^2\right)\Xi$ prescribed by the causality bound~\eqref{eq:causality}. 
Our local vortex resembles the baryonic vortex under a finite isospin chemical potential $\mu_I$ together with $\mu$, pioneered in Refs.~\cite{Qiu:2024zpg,Hamada:2025inf,Hamada:2026uec}. 
The $\mu_I$ would bring in a term $f_{\pi}^{2}\left[\left(1-A_\phi\right)^2-\mu_I^{2}\right]\left(\pi_{1}^{2}+\pi_{2}^{2}\right)/2$ among $\mathcal{E}$, which induces the charged pion condensate and leads further to ANO vortices under a magnetic field~\cite{Adhikari:2015wva,Adhikari:2018fwm,Adhikari:2022cks,Gronli:2022cri}.  
For the local vortex that we study, $f_\pi^2\left(\rho^{-2}-\Omega^2\right)\left(1-A_\phi\right)^2\left(\pi_{1}^{2}+\pi_{2}^{2}\right)/2$ is the counterpart of the $\mu_I$ term.
Unlike in the case of $\mu_I$, however, the finite size effect constrained by eq.~\eqref{eq:causality} is essential in the energetic competition between the two types of vortex states, as we will discuss later.

On the other hand, the ansatz with $\beta = \phi$ leads to a global vortex with $A_\phi = 0$, understood from the positive definite contribution of $A_\phi$ in eq.~\eqref{eq:2kin}.
In the vicinity of the transverse boundary $\rho = R$, we expect a neutral (rather than charged) pion condensate, {\it i.e.}, $\alpha=\pi/2$ because $f_\pi^2\left(R^{-2}-\Omega^2\right)\left(\sigma^{2}+\pi_{3}^{2}\right)/2$ tends to be minimized. 
On top of the neutral pion, to have a finite baryon number reducing $\mathcal{E}_\mathrm{WZW}$, the winding of charged pions centers around the $z$-axis, {\it i.e.}, $\alpha\left(\rho=0\right)=0$, yielding a vortex-Skyrmion
~\cite{Gudnason:2014hsa,Gudnason:2014jga,Gudnason:2016yix,Nitta:2015tua}, 
see also 
refs.~\cite{Gudnason:2020luj,Gudnason:2020qkd} for the viewpoint of a linking number.
Namely, the boundary condition is set up in the style of a vortex, but the homotopy of the solution is that of Skyrmions; $\pi_{3}\left(S^{3}\right) \simeq \mathbb{Z}$.
To the best of our knowledge, a global vortex in the context of baryonic matter has never been discussed before.

\section{Local Vortices\label{sec:local vortex}}
With the identification $\gamma=\phi$, the profile of the local baryonic
vortex is delineated by $\alpha=\alpha\left(\rho,z\right)$
and $\beta=\beta\left(\rho,z\right)$, on top of the gauge field $A_\phi=A_\phi \left(\rho\right)$.
As mentioned, a finite baryon number is desired to reduce the energy via the WZW
term.
The baryon density is further spelled out with the ansatz applied:
\begin{equation}
j_B^{0}=\frac{1}{4\pi^{2}\rho}\left\{ \frac{\partial}{\partial \rho}\left[\cos^{2}\alpha\left(1-A_\phi\right)\frac{\partial \beta}{\partial z}\right]-\left(\rho\leftrightarrow z\right)\right\} .
\end{equation}
Since there is no singularity of $\beta$ in the bulk, we can switch the orders of $\rho$ and $z$ derivatives of $\beta$ freely.
For a single vortex in a finite size cylinder with the radius $R<1/\Omega$ and a longitudinal length
$L$, the net baryon density then reads
\begin{align}
\int d^{3}xj_B^{0}=\frac{1}{2\pi} & \bigg\{\int dz\cdot\left[\cos^{2}\alpha\left(1-A_\phi\right)\frac{\partial \beta}{\partial z}\right]\bigg|_{\rho=0}^{\rho=R}\nonumber \\
 & -\int d\rho\cdot\left[\cos^{2}\alpha\left(1-A_\phi\right)\frac{\partial \beta}{\partial \rho}\right]\bigg|_{z=-L/2}^{z=L/2}\bigg\}.\label{eq:j0}
\end{align}
A configuration with nonvanishing eq.~\eqref{eq:j0} is realized by
longitudinal periodicity:
\begin{equation}
\beta\left(\rho,z+d\right)=\beta\left(\rho,z\right)+2\pi,\quad\alpha\left(\rho,z+d\right)=\alpha\left(\rho,z\right).
\end{equation}
In this way, the total length $L=Nd$ is divided into $N$ identical
units of period $d$. Over each distance $d$, $\beta$ winds $2\pi$,
in analogy to the neutral pion domain wall, also known as the chiral soliton lattice. Meanwhile, $U$ returns to its own value whilst $A_\phi$ is independent
of $z$. Hence the second line of eq.~\eqref{eq:j0} has null contribution.
In contrast, among the first line, given that 
\begin{equation}
A_\phi\left(0\right)=0,\quad A_\phi\left(R\right)=1,\label{eq:bca}
\end{equation}
the pure gauge vanishes $j_B^{0}$ at the boundary $\rho=R$. 
Eventually, 
\begin{equation}
\int d^3x j_B^0 = N\cdot2\pi \int_{0}^{R}\int_{0}^{d}\rho d\rho dz j_B^{0} = N
\label{eq:N}
\end{equation}
must come solely from the winding of $\beta$ at $\rho=0$, {\it i.e.},
\begin{align}
\alpha\left(0,z\right) & =0,\quad\alpha\left(R,z\right)=\pi/2;\label{eq:bcalpha}\\
\beta\left(0,0\right) & =0,\quad\beta\left(0,d\right)=2\pi.\label{eq:bcbeta}
\end{align}
As long as eqs.~\eqref{eq:bca}, \eqref{eq:bcalpha} and~\eqref{eq:bcbeta}
are satisfied, the vortex carries $N$ baryon number.
To obtain a vortex-Skyrmion state, hence, we adopt these equations as the boundary condition for the phase functions $\alpha$ and $\beta$, and gauge field.

Since $d$ is also a parameter to be physically determined, the problem involves a bilevel optimization; a soliton solution is obtained from the minimization of the string tension defined as
\begin{equation}
T\equiv\frac{1}{L}\int d^{3}x\mathcal{E}=\frac{1}{d}\left[-\mu+2\pi\int_{0}^{R}\int_{0}^{d}\rho d\rho dz\left(\mathcal{E}_{\text{kin}}+\mathcal{E}_{\text{EM}}\right)\right],
\label{eq:tension}
\end{equation}
which equals the total energy divided by a fixed total length $L$, but reduces the volume integral to that within one longitudinal period.
The first term $-\mu/d$ corresponds to the contribution from the WZW term.
The minimization of $T$ is done by two steps.
First, at a certain value of $d$, we solve $\alpha$, $\beta$ and
$A_\phi$ from $\delta\mathcal{E}|_{d} = 0$.
Next, we apply their solutions
to compute eq.~\eqref{eq:tension} numerically and repeat doing so
for different input values of $d$, 
yielding the period that minimizes $T$, {\it i.e.},
\begin{equation}
d_0=\text{argmin}\,T(d),
\end{equation}
which is the period for the physical solution. In other words, if $T(d)$ has no minimum, it means that a vortex solution cannot be found or proves unstable.

From now on, we consider a simplified ansatz valid in the chiral limit,
assuming separation of variables 
\begin{equation}
\alpha=\alpha\left(\rho\right),\quad\beta=\beta\left(z\right),\label{eq:separates}
\end{equation}
as done in our preceding work~\cite{Qiu:2024zpg}. We shall disclaim
that such a simplified configuration cannot guarantee a minimum energy,
but it could feature lower energy than a pionic vortex or even homogeneous
pion condensate. 
If the latter occurs, a magnetic field is spontaneously
generated along $\boldsymbol{\Omega}$. The specific form of energy
density with eq.~\eqref{eq:separates} applied is 
\begin{align}
\mathcal{E}_L= & \frac{f_{\pi}^{2}}{2}\left[\left(\frac{\partial \alpha}{\partial \rho}\right)^{2}+\cos^{2}\alpha\left(\frac{\partial \beta}{\partial z}\right)^{2}+\left(\frac{1}{\rho^{2}}-\Omega^{2}\right)\left(1-A_\phi\right)^{2}\sin^{2}\alpha\right] \nonumber
\\
 & +\frac{1}{2e^{2}}\left(\frac{1}{\rho^{2}}-\Omega^{2}\right)\left(\frac{\partial A_\phi}{\partial \rho}\right)^{2},
 \label{eq:Elocal}
\end{align}
with which $\delta\mathcal{E}_L|_{d}=0$ leads to the following equations of motion (EOMs):
\begin{equation}
\frac{\partial^{2}\beta}{\partial z^{2}}=0,\label{eq:eombeta}
\end{equation}
\begin{equation}
\frac{\partial}{\partial\rho}\left(\rho\frac{\partial\alpha}{\partial\rho}\right)=\frac{1}{2}\rho\sin2\alpha\left[\left(\frac{1}{\rho^{2}}-\Omega^{2}\right)\left(1-A_\phi\right)^{2}-\left(\frac{\partial\beta}{\partial z}\right)^{2}\right],
\label{eq:eomalpha}
\end{equation}
\begin{equation}
\frac{\partial}{\partial\rho}\left[\rho\left(\frac{1}{\rho^{2}}-\Omega^{2}\right)\frac{\partial A_\phi}{\partial\rho}\right]=-e^{2}f_{\pi}^{2}\rho\left(\frac{1}{\rho^{2}}-\Omega^{2}\right)\left(1-A_\phi\right)\sin^{2}\alpha
\label{eq:eoma}.
\end{equation}
One straight corollary from eq.~\eqref{eq:eombeta} is
\begin{equation}
\beta=\frac{2\pi}{d}z, 
\label{eq:beta}
\end{equation}
in accordance with the boundary condition in eqs.~\eqref{eq:bcbeta}. 
Also we remind that eq.~\eqref{eq:eoma} is nothing but the Maxwell equation
\begin{equation}
\partial_{\mu}F^{\mu\nu}=j_Q^{\nu}=qj_B^{\nu}+j_I^{\nu},
\end{equation}
which features the isospin current 
\begin{equation}
j_I^{\mu}\equiv\frac{\delta}{\delta A_{\mu}}\left(\mathcal{L}-\mathcal{L}_{\text{EM}}\right),
\end{equation}
and the vanishing baryon current $j_B^{\phi}=0$. The WZW term does
not enter EOMs at this stage for a fixed $d$. 
However, it participates
in determining the specific value of $d_0$.
Apart from the $\beta$ which is analytically solved in eq.~\eqref{eq:beta},
$\alpha$ and $A_\phi$ are further solved from the above EOMs~\eqref{eq:eomalpha} and \eqref{eq:eoma}
under the boundary conditions eqs.~\eqref{eq:bcalpha}, \eqref{eq:bcbeta} and \eqref{eq:bca}.

\section{Global Vortices}
\label{sec:global vortex}

For the global vortex with $\beta=\phi$, the baryon density turns out independent of $A_\phi$:
\begin{equation}
j_B^{0}=-\frac{1}{4\pi^{2}\rho}\left[\frac{\partial}{\partial \rho}\left(\cos^{2}\alpha\frac{\partial \gamma}{\partial z}\right)-\left(\rho\leftrightarrow z\right)\right].
\end{equation}
Following the transcript of sec.~\ref{sec:local vortex}, we consider
the global vortex with longitudinal periodicity,
\begin{equation}
\gamma\left(\rho,z+d\right)=\gamma\left(\rho,z\right)+2\pi,\quad\alpha\left(\rho,z+d\right)=\alpha\left(\rho,z\right).
\end{equation}
The integral $\int d^3x j_B^{0}$ consists in the winding of $\gamma$
at $\rho=R$ where the neutral pion condensates. 
Essential boundary
conditions to realize eq.~\eqref{eq:N} take the
form:
\begin{align}
\alpha\left(0,z\right) & =\pi/2,\quad\alpha\left(R,z\right)=0;\label{eq:bcag}\\
\gamma\left(R,0\right) & =2\pi,\quad\gamma\left(R,d\right)=0.\label{eq:bcgamma}
\end{align}
Again we leverage the simplified ansatz with the separation of transverse
and longitudinal variables
\begin{equation}
\alpha=\alpha\left(\rho\right),\quad\gamma=\gamma\left(z\right).
\end{equation}
It leads to the specific form of the energy density
\begin{align}
\mathcal{E}_G= & \frac{f_{\pi}^{2}}{2}\left[\left(\frac{\partial \alpha}{\partial \rho}\right)^{2}+\sin^{2}\alpha\left(\frac{\partial \gamma}{\partial z}\right)^{2}+\left(\frac{1}{\rho^{2}}-\Omega^{2}\right)\left(\cos^{2}\alpha+A_\phi^{2}\sin^{2}\alpha\right)\right]
\nonumber
\\
 & +\frac{1}{2e^{2}}\left(\frac{1}{\rho^{2}}-\Omega^{2}\right)\left(\frac{\partial A_\phi}{\partial \rho}\right)^{2}
 \label{eq:Eglobal}
\end{align}
Unlike in eq.~\eqref{eq:Elocal}, the contribution of $A_\phi$ in the $\mathcal{E}_G$ is positive definite. Thus, we deduce
\begin{equation}
A_\phi= 0,
\end{equation}
which is consistent with intuition since the global vortex is associated with the neutral pion that does not couple to electromagnetic fields (aside from the anomaly).
Other EOMs read
\begin{equation}
\frac{\partial^{2}\gamma}{\partial z^{2}}=0,\label{eq:geombeta}
\end{equation}
\begin{equation}
\frac{\partial}{\partial\rho}\left(\rho\frac{\partial\alpha}{\partial\rho}\right)=\frac{\rho}{2}\sin2\alpha\left[-\left(\frac{1}{\rho^{2}}-\Omega^{2}\right)+\left(\frac{\partial\gamma}{\partial z}\right)^{2}\right],\label{eq:geomalpha}
\end{equation}
Similar to the local vortex, immediately solved is
\begin{equation}
\gamma=-\frac{2\pi}{d}z.
\end{equation}
Then, essentially the problem
boils down to 1D, {\it i.e.}, eq.~\eqref{eq:geomalpha} for $\alpha(\rho)$, 
which is numerically solved in a parallel manner to that of the local vortex.

\section{Numerical Results}
\label{sec:phase}

\subsection{Parameter Ranges}
We discuss the parameter ranges and the setup for our numerical analysis. 
Throughout this study, we adopt a constant pion decay constant $f_{\pi}=93\text{ MeV}$ to obtain quantitative results.
Regarding the system size, in a realistic bulk system, the transverse radius $R$ effectively characterizes the finite size of the medium (e.g., a fireball) or the effective radius in the cylindrical Wigner-Seitz cell approximation for a vortex lattice. Thus, we take $R\sim 1-10\,\text{fm}$ in our computation.
Besides, the causality constraint~\eqref{eq:causality} stipulates the upper bound of $R$ for a given rotation $\Omega$.

In setting the parameter ranges for $\mu$ and $\Omega$, we select values that represent the physical boundaries of rotating nuclear matter to provide a comprehensive theoretical map of the competition between local and global vortices. The high-density region around $\mu \sim 1--2$ GeV is examined as a theoretical limit where baryonic effects are maximized. Furthermore, while the rotation in realistic baryon-rich objects like neutron stars is negligibly small at the $f_\pi$ scale, the case of $\Omega \to 0$ serves as a crucial benchmark to isolate the role of rotation by supplying the static high-density limit for comparison.

We also comment on a restriction about the baryon chemical potential $\mu$.
In our numerical computation, we minimize the string tension $T(d)$ in eq.~\eqref{eq:tension}, where a local minimum is determined through the competition between the contribution from the WZW term $-\mu/d$ and the others $\int d^3x (\mathcal{E}_\text{kin}+\mathcal{E}_\text{EM})/d$.
We have numerically verified that for a sufficiently small $\mu$, the string tension $T(d)$ monotonically decreases with an increasing $d$, having no minimum. 
It means that the WZW contribution is too small to overcome the (kinetic and electromagnetic) energy cost by forming the inhomogeneous vortex. 
Besides, there exists no minimum of $T(d)$ for large $\mu$ too. But in this case, conversely, $T(d)$ is a monotonically increasing function of $d$. As mentioned, this is because the present framework is within $\mathcal{O}(p^2/\Lambda_\chi^2)$, where $-\mu/d$ can blow up.  
Such an artifact would be remedied if the $\mathcal{O}\left(p^{4}/\Lambda_\chi^4\right)$ contributions are taken into account, as discussed in ref.~\cite{Qiu:2024zpg}.
With these clarifications, in the following analysis, we restrict ourselves to the specific range of $\mu$ where local minima of $T(d)$ exist.

\subsection{Zero Rotation}
\label{subsec:zero}

As a benchmark to understand the consequences of rotation, we first analyze the limit of $\Omega = 0$. In this static regime, although the causality bound $R < \Omega^{-1}$ formally vanishes, we continue to adopt a finite $R$ as the characteristic scale defined in sec.~\ref{sec:phase}. This allows us to handle the logarithmic gradient energy of global vortices through the finite-volume regularization ({\it i.e.}, the Wigner-Seitz cell approximation) and establish a baseline for the string tension before introducing rotational effects.

\begin{figure}[!htb] 
\minipage{0.33\textwidth}   
\includegraphics[width=\linewidth]{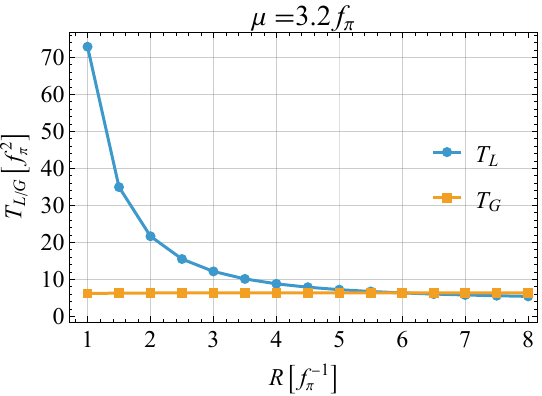}
\endminipage\hfill 
\minipage{0.33\textwidth}   
\includegraphics[width=\linewidth]{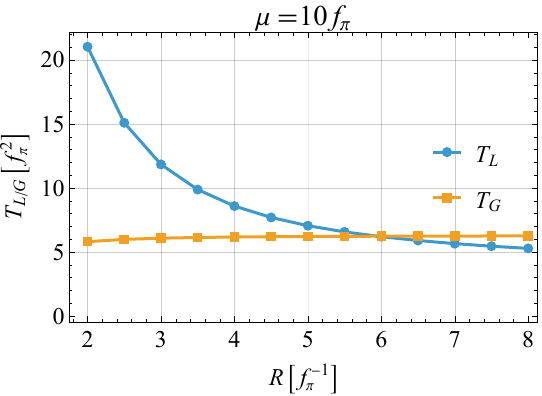}
\endminipage\hfill 
\minipage{0.33\textwidth}   
\includegraphics[width=\linewidth]{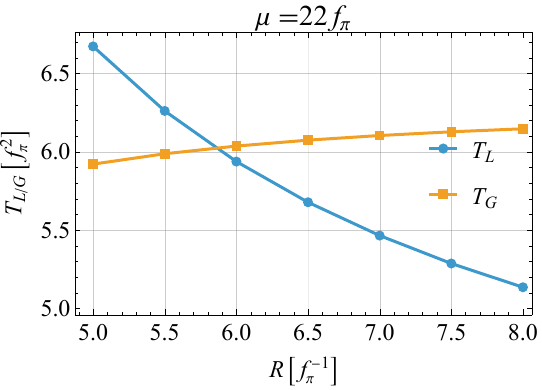}
\endminipage\hfill 
\label{fig:zero}
\caption{The string tension of local/global vortex solutions at zero rotation $\Omega=0$ and three typical values of the baryon chemical potential $\mu$, applied to QGP or neutron stars. For $R$ smaller than the presented ranges, $T(d)$ increases monotonically with an increasing $d$, yielding $d_0\rightarrow0$, which means there is no stable vortex solution with finite size.}
\label{fig:omega0}
\end{figure}

In fig.~\ref{fig:omega0}, we evaluate and plot the minimized string tensions $T_{L/G}$ at varied values of $R$ within the valid regions where stable finite-size solutions exist. The intersection of $T_G=T_L\equiv T_c$ on the $T$\text{--}$R$ plane signifies an energetic competition between the two types of vortex states.
At the transition, the longitudinal periods of local and global vortices differ from each other. Specifically, for $\mu\in(940\text{ MeV},\,2\text{ GeV})$ we find $d_0^L\in(67,167)\text{ fm}$ and $d_0^G\in(91,212)\text{ fm}$.
Another crucial observation is that the transition between the two types of vortex states is insensitive to the value of $\mu$ (at least within our parameter range):
The transition commonly occurs near $R_c\simeq6f_\pi^{-1}=12.7 \text{ fm}$.\footnote{To be more precise, the intersections of $T_L$ and $T_G$ occur at $R_c=5.96$ for $\mu=3.2f_\pi$, $R_c=5.95$ for $\mu=10f_\pi$, and $R_c=5.86$ for $\mu=22f_\pi$.}
We have confirmed such insensitivity of $\mu$ even for finite rotation.

\subsection{Finite Rotation}

In a similar way, we compute $T_L(\Omega)$ and $T_G(\Omega)$, the string tensions of global and local vortices under finite rotation with $\Omega \neq 0$. 
As stated in sec.~\ref{subsec:zero}, a change of $\mu$ (at least under 2 GeV), would not influence the critical phenomenon significantly.
Hence, selecting one typical value of $\mu$ is sufficient for the study in the current subsection.
In such a circumstance, we focus on the case with $\mu=300\text{ MeV}\simeq3.2f_\pi$, which is a typical scale of baryon chemical potential in the hadronization procedure. 
We have confirmed numerically that $T_L(\Omega)$ and $T_G(\Omega)$ have an intersection of string tensions within the window 
\begin{equation}
  R\in(5.97f_\pi^{-1},\,6.17f_\pi^{-1})\equiv(R_1,\,R_2),
  \label{eq:window}
\end{equation}
indicating a transition at the critical angular velocity $\Omega_c$, which is defined as 
\begin{equation}
T_L\big|_{\Omega=\Omega_c}
=T_G\big|_{\Omega=\Omega_c}
\equiv T_c.
\end{equation}
On the other hand, we have observed
\begin{align}
    & R<R_1:\quad T_G\left(\Omega\right)<T_L\left(\Omega\right),\\
    & R> R_2:\quad T_G\left(\Omega\right)>T_L\left(\Omega\right),
\end{align}
implying the absence of such a transition outside the window~\eqref{eq:window}.
An explanation of this is that the larger system size ($R > R_2$) would aggravate the logarithmic divergence in energy, exclusive to the global vortex but not the local vortex.
Conversely, for a smaller system size ($R < R_1$), the local vortex is strongly disfavored because accommodating the boundary condition for the gauge field within a small radius drastically increases the electromagnetic energy cost.

\begin{figure}[!htb] 
\minipage{0.33\textwidth}   
\includegraphics[width=\linewidth]{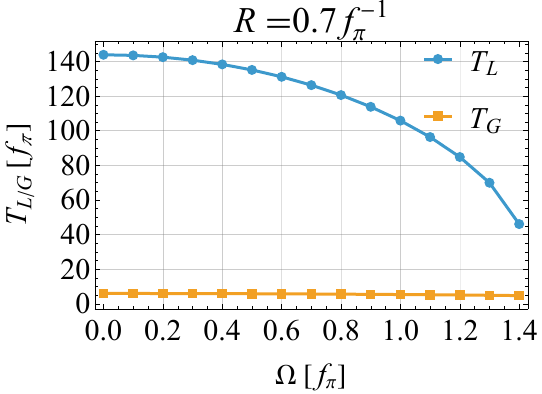}
\endminipage\hfill 
\minipage{0.33\textwidth}   
\includegraphics[width=\linewidth]{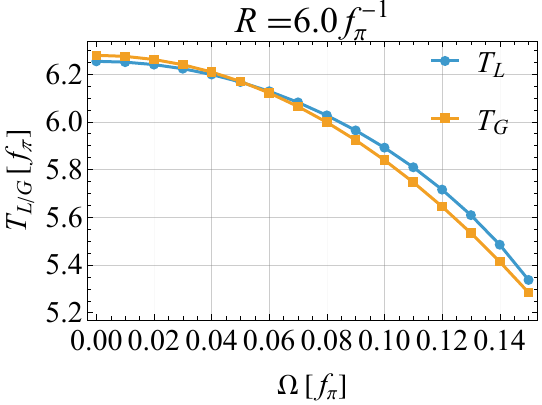}
\endminipage\hfill 
\minipage{0.33\textwidth}   
\includegraphics[width=\linewidth]{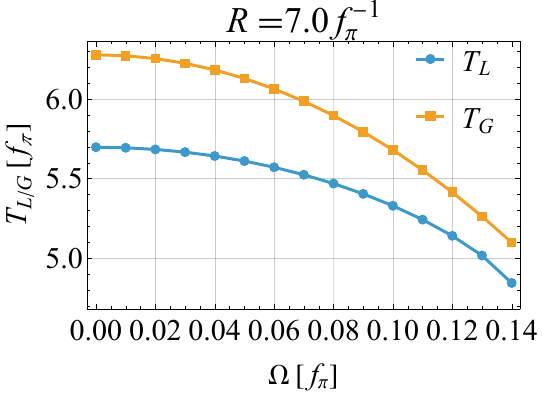}
\endminipage\hfill
\caption{The string tension of local/global vortex solutions as a function of $\Omega$ at different system radii $R$ and common baryon chemical potential $\mu=3.2f_\pi\simeq300\text{ MeV}$.}
\label{fig:typical}
\end{figure}

We illustrate $T_{L/G}\left(\Omega\right)$ at three different typical $R$ in fig.~\ref{fig:typical}.
For $R<R_1$, we choose the nucleon radius around $1\text{ fm}\simeq 0.7f_\pi$ as a typical value of the cutoff.
On the other hand for $R>R_2$ we select a value ten times larger, $10\text{ fm}\simeq 7f_\pi$, 
representing a macroscopic transverse scale.

\begin{figure}[!htb] 
\minipage{0.33\textwidth}   
\includegraphics[width=\linewidth]{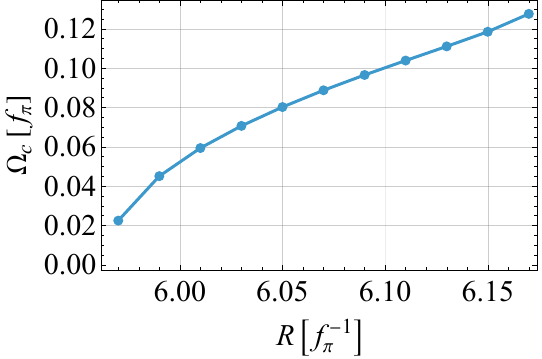}
\endminipage\hfill 
\minipage{0.34\textwidth}  
\includegraphics[width=\linewidth]{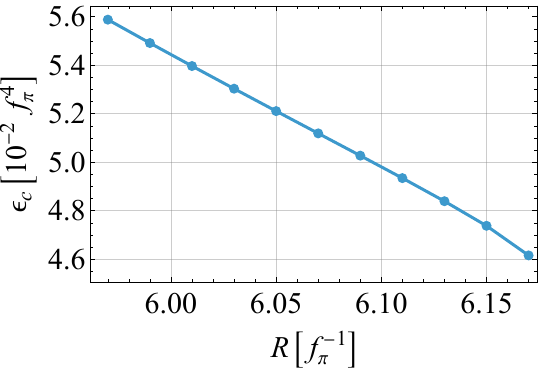}
\endminipage\hfill
\minipage{0.33\textwidth}  
\includegraphics[width=\linewidth]{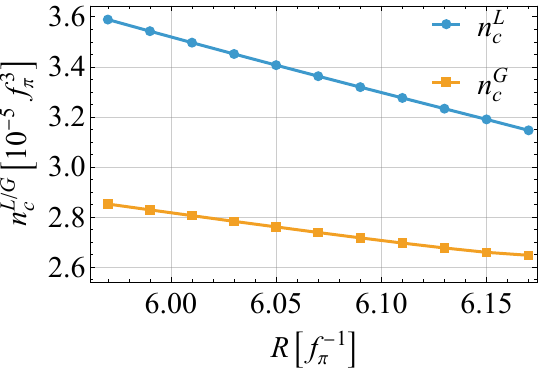}
\endminipage\hfill
\caption{Critical angular velocity, energy density, and baryon density at $\mu=300\text{ MeV}$.}
\label{fig:critical}
\end{figure}

Now we focus on the case when $R\in\left(R_1,R_2\right)$. At the critical angular velocity $\Omega_c$, although both types of vortices feature a common string tension $T_c$, or the energy density
\begin{equation}
    \epsilon_c\equiv \frac{T_c}{\pi R^2},
\end{equation}
their longitudinal period differ $d_0^L\neq d_0^G$, so do their number densities 
\begin{equation}
n_c^{L/G}\equiv  \frac{1}{\pi R^2 d_0^{L/G}}.
\end{equation}
Since within one longitudinal period, a vortex carries one baryon number, the $n_c^{L/G}$ defined above is exactly the baryon density associated with the vortex. 
In fig.~\ref{fig:critical}, we illustrate the shift of the critical point depending on $R$ via the physical quantities $\Omega_c$, $\epsilon_c$, and $n_c^{L/G}$.

To glean physical insights from these results, we note that the critical angular velocity $\Omega_c\sim0.01f_\pi$\text{--}$0.1f_\pi$ [fig.~\ref{fig:critical} (left)] aligns well with the typical scales of heavy-ion collisions ($10$\text{--}$100$ MeV). 
Quantitative values of the effective energy/baryon densities [fig.~\ref{fig:critical} (mid/right)] are subject to the limitations of leading-order ChPT lacking $\mathcal{O}(p^4)$ terms. However, we emphasize that the qualitative transition remains robust, driven by the universal competition between the logarithmic kinetic energy portion of the global vortex and the electromagnetic energy cost of the local vortex, as discussed above.

Importantly, we infer the message that at a realistic $R\ll R_1$, the global vortex should be energetically favored over the local vortex. 
The possibility of such a global vortex has been largely overlooked previously due to the logarithmic energy divergence in infinite systems, whereas our study indicates it should play a crucial role as a viable topological excitation in dense hadronic matter under rotation.

\section{Conclusion and Outlook}
\label{sec:summary} 

In this work, for the first time, we have discussed a global vortex that is relevant to topological excitations of baryonic matter under rotation. 
The rotation does not distinguish between neutral and charged pions, so either can form a condensate, leading further to a global or local vortex, respectively. 
Global vortices were largely overlooked before due to the logarithmic divergence in energy. However, the rotation comes with the causality bound that constrains the system size, making a finite-energy global vortex possible. Furthermore, we demonstrate that both types of vortex can be stabilized as topological excitations by the WZW term, given a finite baryon number carried by the vortex and coupling to the baryon chemical potential. 
Such vortex solutions are technically vortex-Skyrmions of baryonic nature, carrying a Skyrmion-type homotopy $\pi_3(S^3)$ but without a Skyrme term, and therefore model independent.

Indeed, we have found that the energetic competition between global and local vortex configurations, whether relevant or not, is highly sensitive to the transverse (perpendicular to the vortex line) scale $R$. 
For either zero rotation or maximum rotation bounded by causality, the transition can be witnessed only within a small window $R_0\pm\Delta R$ centering around $R_0\approx 6 f_\pi^{-1}$ with $\Delta R\sim 0.5 \%R_0$. 
Such a scale $R_0$ shall be compared with the lattice period of the rotating vortices in realistic physics contexts. 
For a smaller system size, the global vortex is energetically favored, while for a larger system size, the local vortex dominates. This indicates that a previously overlooked metastable topological defect can emerge under rapid rotation when the system size is sufficiently constrained.

The workflow we developed applies to broader theories associated with vortices, such as the Abelian-Higgs model, among others. 
Generally speaking, the underlying mechanism consists in the fact that when the system size is comparable with the penetration depth, the logarithmic divergence in the energy of the global vortex is remedied. Thus, global and local vortices can have energies on a similar scale.
For even smaller system size, the local vortex becomes disfavored due to the large electromagnetic energy, while for large system size, the global vortex bears divergent kinetic energy.

Based on a common ChPT framework, we have several outlooks:
(1) Extend the theory to $\mathcal{O}\left(p^{4}/\Lambda_\chi^4\right)$, including especially the Skyrme term. As noted in sec.~\ref{sec:phase}, the absence of such stabilizing terms in the leading-order theory prevents the soliton from compactifying along the longitudinal direction, resulting in an unnaturally large period $d_0$ and dilute densities. Incorporating the Skyrme term is essential to fully resolve the longitudinal structure of the vortex-Skyrmion and allow for a direct quantitative comparison with the bulk densities of realistic nuclear matter.
(2) Generalize the study from a single vortex to a vortex lattice, which will feature a lattice constant replacing the $R$, providing a more realistic description of rotating macroscopic media.
(3) Introduce finite-temperature effects, which will be a critical step toward understanding the dynamical formation and survivability of these topological defects during the expansion of the QGP fireball.
These prospects will connect our theoretical baseline with phenomenology in heavy-ion collisions and neutron stars.

\begin{acknowledgments}
 This work is supported in part by JSPS Grant-in-Aid for Scientific Research KAKENHI Grant JP24K17052 (K.\,M.), No. JP22H01221 and JP23K22492 (M.~N. and Z.~Q.).
The work of M.~N. is supported in part by the WPI program ``Sustainability with Knotted Chiral Meta Matter (WPI-SKCM$^2$)'' at Hiroshima University.
 
\end{acknowledgments}

\bibliographystyle{JHEP}
\bibliography{rotation,soliton}

\end{document}